\documentclass{article}

\usepackage{arxiv}

\usepackage[utf8]{inputenc} 
\usepackage[T1]{fontenc}    
\usepackage{lmodern}        
\usepackage{hyperref}       
\usepackage{url}            
\usepackage{booktabs}       
\usepackage{amsfonts}       
\usepackage{nicefrac}       
\usepackage{microtype}      
\usepackage{graphicx}
\usepackage{bbm}

\title{A Bayesian Hierarchical Time Series Model for Reconstructing
Hydroclimate from Multiple Proxies}

\author{
    Niamh Cahill
   \\
    Department of Mathematics and Statistics \\
    Maynooth University \\
  Kildare, Ireland \\
  \texttt{\href{mailto:niamh.cahill@mu.ie}{\nolinkurl{niamh.cahill@mu.ie}}} \\
   \And
    Jacky Croke
   \\
    School of Geography \\
    University College Dublin \\
  Dublin, Ireland \\
  \texttt{} \\
   \And
    Micheline Campbell
   \\
    School of Biological, Earth and Environmental Sciences \\
    University of New South Wales \\
  Sydney, Australia \\
  \texttt{} \\
   \And
    Kate Hughes
   \\
    Catchment Connections \\
  Brisbane, Australia \\
  \texttt{} \\
   \And
    John Vitkovsky
   \\
    Department of Environment and Science \\
    Queensland Government \\
  Australia \\
  \texttt{} \\
   \And
    Jack Eaton Kilgallen
   \\
    Hamilton Institute \\
    Maynooth University \\
  Kildare, Ireland \\
  \texttt{} \\
   \And
    Andrew Parnell
   \\
    Hamilton Institute \\
    Maynooth University \\
  Kildare, Ireland \\
  \texttt{} \\
  }

\newlength{\cslhangindent}
\newlength{\csllabelwidth}
\newlength{\cslentryspacingunit} 
  {}%
  {\par}
\newenvironment{CSLReferences}[2] 
 {
  \setlength{\parindent}{0pt}
  \ifodd #1
  \let\oldpar\par
  \def\par{\hangindent=\cslhangindent\oldpar}
  \fi
  \setlength{\parskip}{#2\cslentryspacingunit}
 }%
 {}
\usepackage{calc}

\begin{document}
\maketitle

\begin{abstract}
We propose a Bayesian hierarchical model which produces probabilistic reconstructions of hydroclimatic variability in Queensland Australia. The model  provides a standardised approach to hydroclimate reconstruction using multiple palaeoclimate proxy records derived from natural archives such as speleothems, ice cores and tree rings. The method combines time-series modelling with inverse prediction to quantify the relationships between a given hydroclimate index and relevant proxies over an instrumental period and subsequently reconstruct the hydroclimate back through time. We present case studies for Brisbane and Fitzroy catchments focusing on two hydroclimate indices, the Rainfall Index (RFI) and the Standardised Precipitation-Evapotranspiration Index (SPEI). The probabilistic nature of the reconstructions allows us to estimate the probability that a hydroclimate index  in any reconstruction year was lower (higher) than the minimum (maximum) value observed over the instrumental period. In Brisbane, the RFI is unlikely (probabilities < 20\%) to have exhibited extremes beyond the minimum/maximum values observed between 1889 and 2017.  However, in   Fitzroy there are several years during the reconstruction period where the RFI is likely  (>50\% probability) to have exhibited behaviour beyond the minimum/maximum of what  has been observed. For SPEI, the probability of observing such extremes since the end of the   instrumental period in 1889 doesn’t exceed 50\% in any reconstruction year in Brisbane or Fitzroy.
\end{abstract}


\newpage
\hypertarget{introduction}{%
\section{Introduction}\label{introduction}}

The Australian continent experiences hydroclimatic variability that is
regarded as extreme compared to the rest of the world (Dixon et al.
2019; Dowdy et al. 2019). Australia is particularly vulnerable to
relatively small changes in rainfall; floods and droughts are common
(O'Donnell et al. 2021). Past trends in precipitation showing rainfall
increases in many tropical areas and rainfall decreases in many
temperate areas are projected to continue and become more extreme in the
future (Head et al. 2014). Increases in the frequency and intensity of
extremes will create new pressures and vulnerabilities with the impacts
varying geographically and socioeconomically. For example, following
repeated Queensland floods, some insurance companies have increased
premiums and become less willing to insure houses on floodplains
(Suncorp Group 2013). However, considerable uncertainty remains in
relation to how observational extremes (minimum/maximum measurements)
fit within the context of past variability, making it difficult to
attribute more recent observed variation to natural variability or
anthropogenic climate forcing (B. I. Cook et al. 2016; Kiem et al.
2020).

To date, hydroclimatic risk in Australia has been assessed using
primarily historical instrumental records of rain, evaporation, and
streamflow. These instrumental records typically only exist for about
the last 100 years at best. Such short records limit the calculation of
robust statistics around the baseline risk of extreme events (Tingstad,
Groves, and Lempert 2014; Armstrong, Kiem, and Vance 2020) with
potentially catastrophic economic and social consequences. When longer
records are included, more low-probability, higher-impact events are
identified. For example, under climate boundary conditions similar to
present, megadroughts lasting 10-40 years are evident in a 1000-year
eastern Australian annual rainfall reconstruction from the Law Dome (LD)
ice core record (Vance 2012). It is now generally accepted that the
instrumental records do not cover the full range of hydroclimatic
variability and any risk assessments based on them are likely to
underestimate, or at least misinterpret, the frequency, duration,
magnitude and timing of wet and dry periods (Ho, Kiem, and Verdon-Kidd
2015; Tozer et al. 2016).

Proxy records provide indirect estimates of past local or regional
hydroclimate, derived from natural archives such as sediment cores,
speleothems, ice cores and tree rings (Croke et al. 2021). Access to
palaeoclimate proxy databases offers the scientific community an
opportunity to extend instrumental records back in time and better
elucidate natural climate variability. Improving our understanding of
such variability requires making use of statistical methods to
understand relationships between palaeoclimate proxies and direct
measurements of the hydroclimate over an instrumental time period. Armed
with an understanding of the relationship over the instrumental period,
we can extrapolate back through time to estimate and quantify
uncertainty in a reconstruction period. Reconstructions of climate
beyond the period of direct measurements have been performed in numerous
studies that combine appropriate statistical methods and information
from proxy records (Michel et al. 2020; Cahill et al. 2016; Hernández et
al. 2020; Parnell et al. 2015).

Our paper provides details on a new approach to reconstructing the
hydroclimate of selected catchments in Queensland, one of Australia's
most climatically extreme states, using a set of hydroclimate indices
and a newly-available palaeoclimate proxy database that has been
compiled for the Australasian region (Croke et al. 2021). Croke et al.
(2021) identified which of the 396 proxy records included in the
database best correlate with a range of hydroclimate variables for
Queensland catchments. Here, the application of this database is
expanded by presenting a Bayesian time-series, multi-proxy, inverse
modelling approach to produce reconstructions of hydroclimate indices at
the catchment scale in Queensland. 

The proposed framework establishes
(1) the time-series characteristics of the hydroclimate, based on direct
measurements between 1989 and 2017 (referred to as the instrumental
period) and (2) the relationship between palaeoclimate proxies and
the hydroclimate within the instrumental period. We use a state-space model
with a first order autoregressive (AR) process to capture the
variation in the hydroclimate over time and a quadratic regression to model
proxy variation as a function of the hydroclimate. The past (unobserved)
hydroclimate is subsequently reconstructed based on combining time
series back-projection with inverse regression modelling (e.g., Cahill
et al. 2016). The advantages of the approach are, firstly, it provides a standard method that can be applied at any and all locations for which the relevant data are available. Secondly,  multiple proxies are used to inform the hydroclimate reconstruction, which means that strength is drawn from multiple data sources. Thirdly, the choice of what proxies to include is aided by a filtering approach which aims to remove proxy records that have poor modern analogues and finally, using a Bayesian framework holds a major advantage over
other traditional methods (e.g., principal component regression, partial
least squares and elastic net) used in this context (e.g., Luterbacher
et al. 2001; E. R. Cook, D'Arrigo, and Mann 2002; Michel et al. 2020),
as it is conceptually simpler to deal with missing data and to build a
complex model which quantifies the relationship and uncertainties
between multiple proxies and a climate variable.

\hypertarget{data-and-study-area}{%
\section{Data and Study Area}\label{data-and-study-area}}

\label{sec:data}

The proxy and hydroclimate data used in this study was obtained from the
PalaeoWISE (Palaeoclimate Data for Water Industry and Security Planning)
database which comprises 396 quality-assured proxy records, their
metadata, and their relation to Queensland catchment hydroclimate (Croke
et al. 2021). Proxy records in the database are derived from 11
different archive types (e.g.~corals, tree rings, sediments,
speleothems) providing both high- (\(\leq\) 1 year) and low-
(\textgreater1 year) resolution. Further details on the number and range
of proxy archives are presented in Croke et al. (2021).

The selected hydroclimate indices include average annual temperature,
average annual rainfall, average annual evapotranspiration, Standardised
Precipitation Index (SPI) (McKee, Doesken, and Kleist 1993; Adams 2017),
Standardised Precipitation-Evapotranspiration Index (SPEI) (Adams 2017)
and SPI- and SPEI-derived indices for extreme droughts and floods
(McKee, Doesken, and Kleist 1993). Gridded datasets (cell size = 0.05
degrees, approximately 10 km) of instrumental rainfall, temperature and
evapotranspiration were extracted from the Scientific Information for
Landowners (SILO) database
(\url{https://www.longpaddock.qld.gov.au/silo}) for the period 1889 to
2019 using the July to June water year. SPI and SPEI grids (cell size =
0.05 degrees) were then calculated from instrumental data at timescales
of 12, 24, 36, and 48 months, which are standard accumulation periods
used by hydrologists and climatologists. The relationship between the proxy records and catchment-averaged hydroclimate indices was tested in Croke et al. 2021 using correlation analysis across the whole PalaeoWISE database which found that proxies can have the highest correlations with hydroclimate time series at lags of  anywhere between -5 to +5 years. The most appropriate time lag to use is also included in the database and this information is utilised in this study. Preliminary time series analysis for the instrumental hydroclimate indices was carried out using the fable and fpp3 R packages (O'Hara-Wild et al. 2021; Hyndman, Athanasopoulos, and O'Hara-Wild 2021) and the code for running the analysis is available in
\href{https://github.com/ncahill89/wsp/blob/main/vignettes/TS_analysis.md}{this vignette}.

For the purposes of validating our statistical model we use the
hydroclimate and proxy data associated with 18 catchment areas,
including 8 in the Wet Tropics of Queensland, 9 in South East Queensland
and 1 in Fitzroy (Figure 1).

\begin{figure}[h!]
\includegraphics[width=0.75\linewidth]{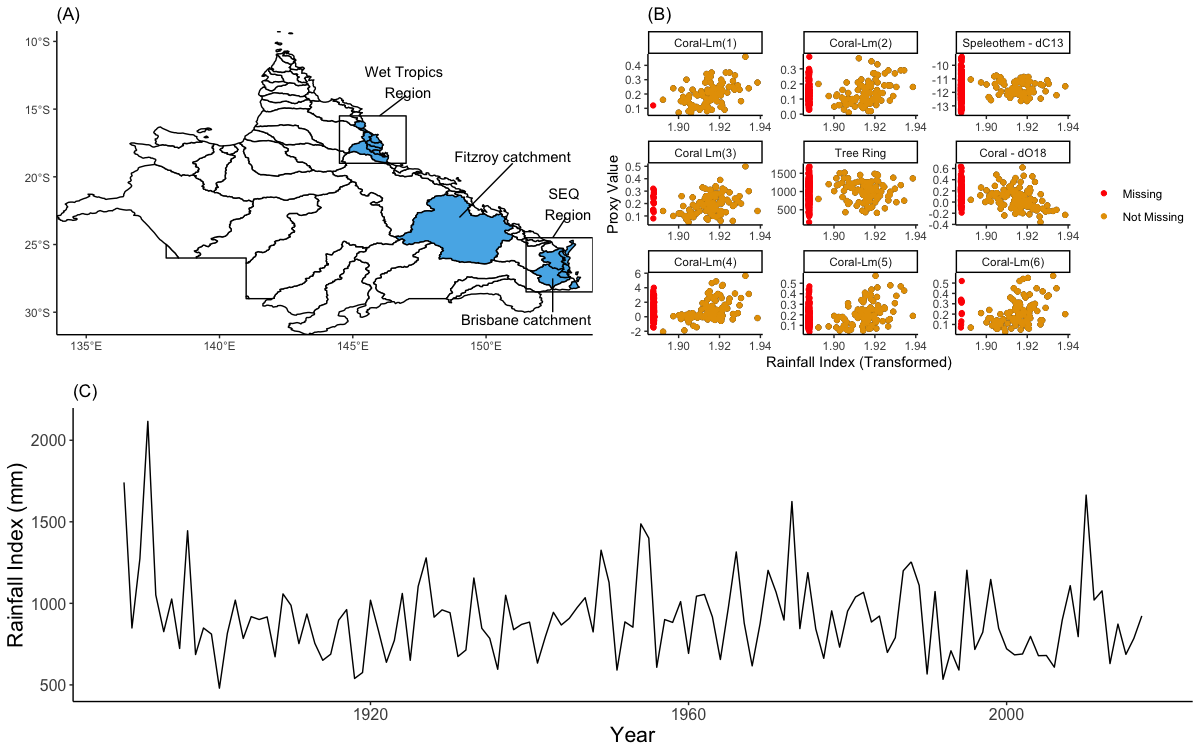} \caption{Map of Queensland and case study catchments and proxy records related to the rainfall index in the Brisbane catchment. (A.) The 18 catchment areas used for the validation are shown in blue. Boxes highlight the Wet Tropics and the South East Queensland (SEQ) regions. Case study catchments Brisbane and Fitzroy are labeled.  (B) Proxy records derived from corals, Tree rings, and Speleothems that have been shown to relate to the rainfall index in Brisbane. The orange points indicate the pairs of proxy and climate data that are available over the instrumental time period. The red points indicate the proxy data that are available over the reconstruction time period (i.e. when instrumental data is missing). (C) The rainfall index (RFI) plotted over the instrumental time period for the Brisbane catchment.}\label{fig:unnamed-chunk-1}
\end{figure}

\newpage

However, when illustrating modelling
results we focus specifically on two case-study catchments that
demonstrate the variability in Queensland's hydroclimate: the
sub-tropical, semi-arid Fitzroy catchment and the sub-tropical Brisbane
catchment (Figure 1). The Fitzroy (\(\sim\) 157 000 km$^2$) catchment is
the largest river catchment in Queensland, draining into the Great
Barrier Reef lagoon, and is the largest cattle-producing region in
Australia. The Brisbane catchment (13,500 km$^2$) is the largest within the
densely populated South East Queensland (SEQ) region (\(\sim\) 27 000
km$^2$), and supports extensive food production. Each catchment has a
number of proxies associated with the various hydroclimate variables as
determined by Croke et al. (2021). For example, Figure 1B shows the
relationship between a number of proxies and average annual rainfall
herein, the rainfall index (RFI), available for the Brisbane catchment.
Figure 1C shows how the RFI has changed over the instrumental time period.

\hypertarget{methods}{%
\section{Methods}\label{methods}}

We propose a Bayesian model which relies on (1) a state-space model with mean modelled by a first order autoregressive (AR) process to capture the variation in a hydroclimate index over time and (2) a quadratic regression model to capture proxy variation across multiple different proxies as a function of the hydroclimate. Quantification of hydroclimate time series dynamics and multiple proxy hydroclimate relationships using calibration data, combined with inverse modelling using historical proxy records, allows us to reconstruct the past hydroclimate drawing on information from multiple proxy input sources. We use simulation-based methods to infer all unknown quantities, including the past (unobserved) hydroclimate, parameters within the model, and uncertainties. The model is structured in a hierarchical framework where data is represented at the data level, the unobserved (latent) quantities are represented at the process level and the parameters and hyper-parameters are represented at the parameter level. The method, described in detail below, can be implemented using the R code and data found in \href{https://github.com/ncahill89/wsp}{this Github repository}.

We will begin by outlining our notation:

\begin{itemize}
\item
  \(I_t\) is a hydroclimate measurement at time \(t\). For the modern
  period (1889 to 2019) these are known. For older periods, these are
  parameters to be estimated.
\item
  \(Y_{ij}\) is the value of proxy \(j\) at time \(t_i\), where \(t_i\)
  is the time point corresponding to observation \(i\).
\item
  We superscript \(I\) and \(Y\) with \(c\) when referring to the
  instrumental time period, such that \(I^c\) and \(Y^c\) correspond to
  the hydroclimate measurements and proxy values that are available in
  the time period 1889 to 2019. We will refer to these data as the
  calibration data.
\item
  We subscript \(I\) and \(Y\) with \(r\) when referring to the
  reconstruction time-period such that \(Y^r\) is the reconstruction
  data and \(I^r\) are parameters to be estimated by the model.
\item
  \(\theta\) is a set of parameters governing the relationship between
  \(Y\) and the expected value of the hydroclimate \(\hat{I}\), where
  for the jth proxy,
  \(\theta_j= \alpha_j,\beta_{j,1} ,\beta_{j,2},\sigma_j\).
  These parameters, which will be defined in the methods section, are informed by the calibration data.
\item
  \(\phi\) is a set of parameters governing the dynamics of how \(I\)
  changes over time, where \(\phi = (\omega, \delta,\rho,\nu,\tau)\). These parameters, which will be defined in the methods section, are informed by the calibration data.
  \end{itemize}

The model has two main components, the data model, and the process
model. Note that all data were standardized to have mean 0 and variance
1 prior to inclusion in the model. 

The rainfall index exhibits non-Gaussian behavior and is transformed using a Box-Cox transformation
prior to analysis. The parameter of the Box-Cox transformation was
chosen via maximum likelihood using the boxcox function from the MASS package (Ripley et al. 2021) in R. The transformation parameter estimate will vary depending on what catchment the rainfall index comes from. For the  Brisbane catchment (Figure 1), the transformation parameter estimate is -0.3. 

\newpage
\hypertarget{data-model}{%
\subsection{Data model}\label{data-model}}

We will assume a univariate state-space model for the observed hydroclimate time series, where we model the data with variability around some level $\gamma$, such that 

\(I_{t} \sim N(\gamma_{t}, \tau^2)\),

The average hydroclimate $\gamma_{t}$ will be modelled as time series process that will be defined in the process modelling section. The variability around $\gamma_{t}$ is captured by $\tau^2$.

The observed proxy data will be linked to their expected values, $\mu_{ij}$, defined in the process modelling section, by assuming that

\(Y_{ij}\sim N(\mu_{ij},\sigma_j^2)\),

where \(\sigma_j^2\) captures the variation of the observations around
the modelled expectation for proxy \(j\).

\hypertarget{process-model}{%
\subsection{Process model}\label{process-model}}

We model the average hydroclimate time series process, $\gamma_{t}$, in one of two ways depending on the dynamics of the series. If there is an apparent trend in the time series then we model  \(\gamma_t\), with a systematic component, which is assumed to be linear, plus a time series component, \(\eta_t\), which captures deviations away from the linear trend, such that, such that   

\(\gamma_t = \omega + \delta t + \eta_t\), 

where the linear trend has intercept \(\omega\) and slope \(\delta\). If there is no apparent trend in the series then 

\(\gamma_t =  \eta_t\).

The time series, \(\eta_t\), is modelled as a first order autoregression AR(1) process, such that

\(\eta_t \sim N(\rho \eta_{t-1},\nu^2)\).

\(\rho\) is the autocorrelation parameter that determines how close consecutive values of \(\eta\) are to each other. We assume a stationary process such that
\(|\rho|<1\), i.e., dependence on the
current state of the series becomes weaker as the distance over time increases. \(\nu^2\) captures the random variation between \(\eta_t\) and \(\rho \eta_{t-1}\).

We let \(\mu_{ij}\), the expected value of the \(i^{th}\) observation of
proxy \(j\), be a quadratic function of the
hydroclimate index at \(t[i,j] \), where \(t[i,j] \) is the time index associated with the
\(i^{th}\) observation of proxy \(j\) for \(j=1,\ldots,M\) proxies and
\(i=1,\ldots,N_j\) time points for proxy \(j\), where

\(\mu_{ij}=\alpha_j +\beta_{j,1}I_{t[i,j]} + \beta_{j,2}I_{t[i,j]}^2 \).

The \(t[i,j] \) indexes are determined based on what lagged version of the hydroclimate time series is most correlated with proxy $j$. For example, if the $i^{th}$ observation of proxy $j$ is observed at time point $t=2$ and if proxy j is best correlated with the series with a lag of -1 then t[i,j] = 1. The data on what lags are used, as determined by Croke et al., 2021 can be found in \href{https://github.com/ncahill89/wsp}{this Github repository}.

\hypertarget{inverse-prediction-via-forward-modeling}{%
\subsection{Inverse prediction via forward
modeling}\label{inverse-prediction-via-forward-modeling}}

Noting that the proxy data, Y, decomposes into $Y = [Y^r, Y^c]$ and the time series, I(t), decomposes into $I(t) = [I^r(t), I^c(t)]$, where the superscripts $r$ and $c$ denote the reconstruction and calibration periods respectively, the overall goal of the approach is to produce the joint posterior
distribution

$p(I^r,\theta,\phi|I^c,Y^c,Y^r) \propto p(Y^c |I^c,\theta) p(I^c|\phi,I^r) p(Y^r |I^r,\theta)p(I^r|\phi)p(\theta)p(\phi).$

The term on the left-hand side of the equation is the posterior
distribution and represents the probability distribution of the
hydroclimate reconstruction given the observed instrumental data. The
terms on the right-hand side represent respectively, the likelihood (the
probability distribution of \(Y^c\) given \(I^c\)), the distribution of
\(Y^r\) given \(I^r\) and the prior distributions of the process and
data model parameters.

\newpage
For the likelihood,

\(p(Y_j^c |\theta,I^c)=\prod_{i=1}^{n_j^c}f_j (y_{ij} |I^c (t[i]),\theta_j),\)

where \(n_j^c\) is the number of observations for proxy \(j\) within the
instrumental time period (1989 to 2019). The set of process and data
model parameters are informed by the calibration data \(I^c\) and
\(Y^c\) contained within this likelihood. In inverse prediction, the
unknown \(I^r\) are added to the set of parameters that need to be
estimated. \(Y_j^r\) are observed and included in the likelihood, where

\(p(Y_j^r |\theta,I^r)=\prod_j(\prod_{i=1}^{n_j^r}f_j (y_{ij} |I^r (t[i]),\theta_j)),\)

and

\(I_t^r |\phi \sim N(\gamma_{t}^r,\tau^2),\)

where \(\theta_j= (\alpha_j,\beta_{j,1} ,\beta_{j,2},\sigma_j)\) and \(\phi=(\omega,\delta,\rho,\nu,\tau)\).

\hypertarget{prior-distributions}{%
\subsection{Prior distributions}\label{prior-distributions}}

The parameters in the model are estimated in a Bayesian framework. Vague
prior distributions are specified for \(\alpha_j\) such that
\(\alpha_j \sim N(0,2^2)\). Laplace priors are placed on the
\(\beta_{k_j}\) coefficients such that
\(\beta_{k_j} \sim Laplace(0,\lambda)\). This is equivalent to a lasso regression, where a Laplace prior places a stronger confidence on zero than a normal prior centered on zero (e.g.,
Tibshirani, 1996). This
prior specification helps to select coefficients for inclusion in the
model, for example if the quadratic term is not strongly supported by the
data it will essentially drop out of the model. We perform selection of
the regularization parameter \(\lambda\) by putting a hyperprior on it,
in this case a half-t prior centered on 0 with scale set at \(5^2\) and 1 degree of freedom. The hydroclimate linear trend parameter,  \(\delta\), is given a normal prior centered on 0 with a variance 1. The autocorrelation parameter of the AR(1) time series process, \(\rho\), is
given a uniform prior between -1 and 1. The standard deviation of the
AR(1) process and the data model standard deviation parameters (\(\nu\)
and \(\sigma_j\)) are given half-t priors centered on 0 with variance
\(2^2\) and 1 degree of freedom. 

\hypertarget{proxy-filtering}{%
\subsection{Proxy filtering}\label{proxy-filtering}}

Prior to running the model outlined above, a proxy filtering procedure
is implemented. Palaeoclimate reconstructions are based on a modern
analogue assumption so that modern relationships, derived from the
overlap of proxy and observed climate data of the instrumental period,
can be used to infer past climate states (e.g., Herbert and Harrison
(2016)). Reconstructions rely on there being some degree of similarity
between the range of proxy values associated with the instrumental and
reconstruction periods. Let \(\bar{y}^c_j\) and \(s^c_j\) be the mean
and standard deviation of the proxy calibration data (i.e., the data
available over the instrumental time period) for proxy \(j\), and let
min($y^r_j$)  and max($y^r_j$) be the minimum and maximum values of the
reconstruction data (i.e., the data available over the reconstruction
time period) for proxy \(j\). We calculate a filtering measure based on
how many standard deviations, \(s^c_j\), the min($y^r_j$) and max($y^r_j$)
are away from \(\bar{y}^c_j\) such that

$F_j = \mbox{max}\bigg(\frac{|\bar{y}^c_j - \mbox{min}(y^r_j)|}{s^c_j},\frac{|\mbox{max}(y^r_j) - \bar{y}^c_j|}{s^c_j}\bigg).$

If \(F_j\) is greater than 3.5 for proxy \(j\), that is, if the minimum
or the maximum proxy value in the reconstruction period is more than 3.5
standard deviations away from the instrumental period mean then we
filter out the proxy record and it will not be used in the
reconstruction. See Figure 2 for an illustration of the filtering
procedure. Proxies are also filtered if they only contain one
observation for the instrumental period. An example of the consequence
of this filtering for reconstructing the RFI in the Brisbane catchment
is that the Coral - Luminescence (1), Coral-\(\delta\)O18 and the Speleothem - \(\delta\)13C
proxies (Figure 1B), were excluded from the reconstruction model.

\begin{figure}[!h]
\includegraphics[width=0.8\linewidth]{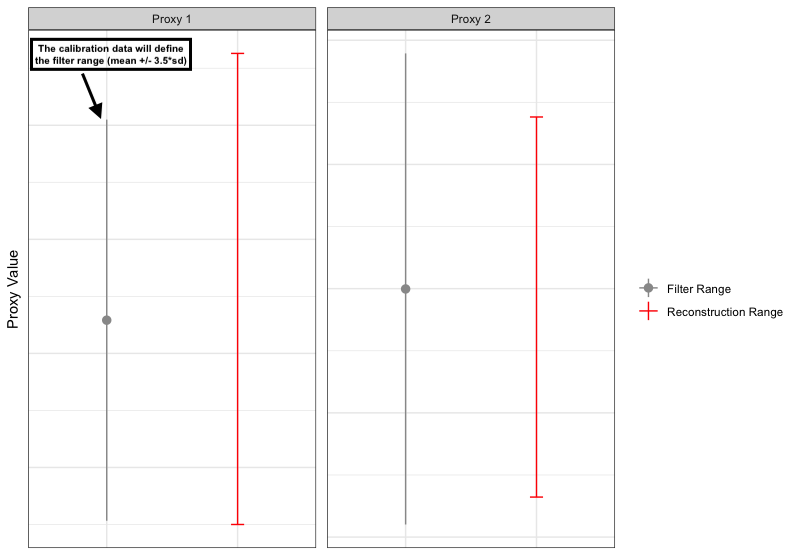} \caption{Illustration of the filtering criteria. The filtering range is shown with the grey vertical bars and is determined by the proxy calibration data. The grey point is the mean of the proxy calibration data. The range of the proxy reconstruction data is illustrated with the red horizontal bar. For Proxy 1, the maximum of the reconstruction data is more than 3.5 standard deviations away from the calibration data mean. Proxy 2 would not be filtered.}\label{fig:unnamed-chunk-2}
\end{figure}

\newpage
\hypertarget{posterior-inference}{%
\subsection{Posterior inference}\label{posterior-inference}}

A Markov chain Monte Carlo (MCMC) algorithm is used to derive samples of
the posterior distributions of the parameters of the model, including
the reconstruction parameter \(I^r\). The MCMC sampling algorithm is
implemented using JAGS (Just Another Gibbs Sampler; Plummer (2003)). To
evaluate the JAGS output we used `rjags,' an R package that offers
cross-platform support from JAGS to the R interface (Plummer, Alexey,
and Denwood 2019). The results include a set of posterior trajectories
for the hydroclimate reconstruction. The median of these results was
taken to be the point estimate and a 95\% credible interval was
calculated using the 2.5\% and 97.5\% percentiles from the posterior
distribution for the reconstructions in each year. The results are based
on a run of 15,000 iterations, of which the first 5,000 are discarded and
the remainder is thinned by saving every 10th sample. The algorithm is
initialized from 3 different starting points yielding three chains
containing 1,000 samples (3000 samples in total) from the posterior
distributions of the model parameters. Decisions on thinning were made to ensure an adequate effective sample size (ESS), such that the ESS would be at least 10\% of the total number of samples. Convergence was checked using the R-hat diagnostics (e.g., Vehtari et al. 2021) and trace plots. R-hat values <1.1 were assumed to be an indicator of convergence. 

\hypertarget{estimating-the-probability-of-exceeding-observed-minimummaximum-values}{%
\subsection{Estimating the probability of exceeding observed
minimum/maximum
values}\label{estimating-the-probability-of-exceeding-observed-minimummaximum-values}}

We make use of the posterior samples of \(I^r(t)\) to estimate the
probability that the hydroclimate index in year $t$ was lower (higher)
than the minimum (maximum) hydroclimate value observed over the
instrumental time period. Let \(I_t^{r(s)}\) be a posterior sample of a
reconstruction in year $t$ and let \(p_t^-\) be the probability that a
hydroclimate index at time \(t\) in the reconstruction period was lower
than the observed minimum (denoted \(I_{min}^c\)), such that

\(p_t^-=(1/N)\sum_{s=1}^N \mathbbm{1}(I_t^{r(s)} < I_{min}^c),\)

where $\mathbbm{1}$ is an indicator function such that $\mathbbm{1}(x) = 1$ if the condition $x$ holds true and 0 otherwise. Similarly, let \(p_t^+\) be the
probability that a hydroclimate index at time t in the reconstruction
period was higher than the observed maximum (denoted \(I_{max}^c\)),
such that

\(p_t^+ = 1-(1/N)\sum_{s=1}^N \mathbbm{1}(I_t^{r(s)} < I_{max}^c).\)

\hypertarget{model-validation}{%
\section{Model Validation}\label{model-validation}}

We assess model performance via a validation exercise which focused on
reconstructing 8 different hydroclimate indices for 18 catchment areas
but leaving out some of the oldest observations from each index. Specifically,
we leave out the 15 oldest data points from each hydroclimate index to
use as test data for an out-of-sample validation. After leaving out
these data we apply the proxy filtering (described in Section 3.5) and
then fit the model to the training data set to obtain reconstruction
estimates and 95\% credible intervals for the test data set. We
calculate coverage of the 95\% credible intervals, mean errors and the root mean-squared error (RMSE) for the test data broken down by hydroclimate index (Table 1).

Note, not all catchment areas will
be represented in the test data associated with each hydroclimate index
and hence the sample sizes in Table 1 vary. In addition, the results are
only presented for records that passed convergence checks. The
mean error provides a measure of the average bias in the model
predictions. Overall, the coverage of the 95\% uncertainty intervals
is 90\%, suggesting that the reconstruction will capture the truth 90\%
of the time. The overall mean error, while close to zero is negative (-0.06) seen in Table 1 which indicates that on average the model is slightly biased towards over-predicting the hydroclimate indices. The RMSE provides a summary of the variation in the prediction errors and can be interpreted as a standard deviation. Table 1 also indicates the
variations in the validation measure when broken down by hydroclimate
index. Coverage ranges from 87\% to 93\%. Ideally all coverage values would be 95\%. The mean errors range from -0.33 to 0.12 with 5 of the indices showing a slight bias towards over prediction and the remainder showing a slight bias towards under prediction. The RMSE, which provides a measure of the spread of the prediction errors, ranges from 0.83 to 1.09.

\hypertarget{case-studies}{%
\section{Case Studies}\label{case-studies}}

This section focuses on applying the model to reconstruct two different
climate indices (1) the rainfall index (RFI) and (2) the Standardised Precipitation Evapotranspiration Index with a 12 month aggregation period (SPEI (12)). We
focus on reconstructing these indices for the Brisbane and Fitzroy
catchment areas. The process model without the trend component was used. Table 2 provides a summary of the proxy data available
for the reconstructions, the number of proxies that were filtered and illustrates that not all proxy records span
the same time period. For example, some of the proxy records used in the
reconstructions will have less of an overlap with the hydroclimate
variables in the instrumental time period relative to others and each of
the proxy records can be of different lengths and temporal resolutions.
The length of a reconstruction will be determined by the length of continuous proxy coverage (in years) once filtering has been applied. The filtering procedure resulted
in 3 proxies being excluded from informing the RFI in Brisbane
(Coral-Luminescence (1), Coral-delta O18, Speleothem-delta C13; Figure
1). For the RFI reconstruction in Fitzroy, 4 proxies were excluded
including 2 coral-luminescence proxies, a tree ring proxy and a
coral-delta-Oxygen-18 proxy. For the SPEI (12) reconstructions,
speleothem-delta-Carbon-13, coral-delta-Oxygen-18, tree ring and
coral-luminescence proxies were excluded for Brisbane.
Coral-luminescence, coral-delta-Oxygen-18 and a tree ring proxy were
excluded for Fitzroy. This filtering limited the proxies here to just
two archives, tree rings and coral luminescence, both of which have
well-established physical mechanisms which describe their relationships
to hydroclimate (Ho, Kiem, and Verdon-Kidd 2015; K et al. 2009). The RFI
and SPEI (12) reconstructions extend back to 1612 CE and 1030 CE for the
Brisbane and Fitzroy catchments respectively (Table 2)

In the next section we present probabilistic reconstructions and
analysis of extremes values relative to observed minimum and maximum
values over the instrumental period for the RFI and the SPEI (12) in
case study catchments Brisbane and Fitzroy. Detailed annual model-based
estimates (plus uncertainty) of the RFI and SPEI for these catchment
areas are available \href{https://github.com/ncahill89/wsp}{here}.
Probabilistic reconstructions and detailed annual estimates (plus
uncertainty) for other hydroclimate variables and catchment areas can be
produced using our \href{https://github.com/ncahill89/wsp}{R code}.

\hypertarget{rainfall-index-reconstruction}{%
\subsection{Rainfall Index
Reconstruction}\label{rainfall-index-reconstruction}}

Figure 3 shows RFI reconstructions (mm) for the Brisbane and Fitzroy
catchments. In the Brisbane catchment, on average the estimated RFI over
the reconstruction period is lower than the average observed RFI over
the instrumental period (a value of 892 mm compared to 926 mm). The
reconstruction period estimates also exhibit less variability than the
observed RFI, with a standard deviation of 158 mm for the reconstruction
period compared to 259 mm for the instrumental period. In the Fitzroy
catchment, the average estimated RFI for the reconstruction period is
667 mm compared to 678 mm for the instrumental period. The
reconstruction period estimates exhibit slightly less variability than
the observed RFI, with a standard deviation of 195 mm compared to 198
mm. It is worth noting that the start of the instrumental period in
these catchments coincided with the timing of some of the largest floods
on record (Bureau of Meteorology 2017).

\begin{figure}[!h]
\includegraphics[width=0.9\linewidth]{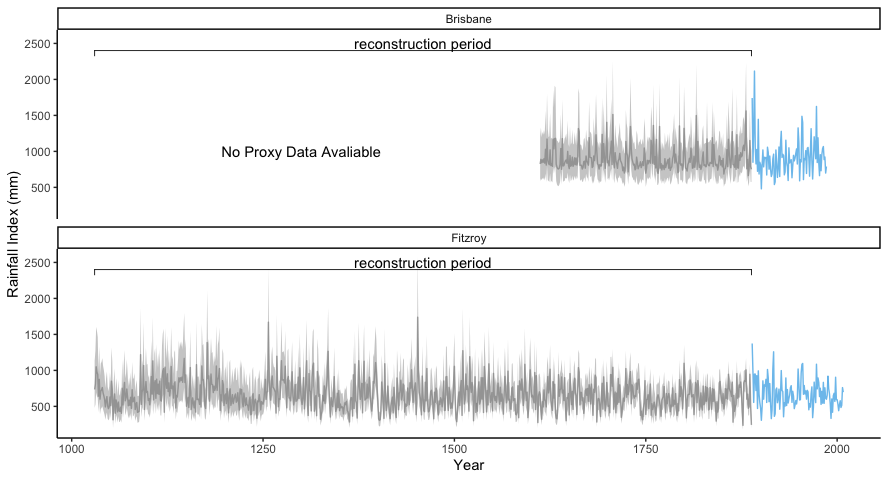} \caption{Rainfall Index (RFI) in mm reconstructions for the Brisbane and Fitzroy catchment areas. The solid blue line represents the observed instrumental data. The solid grey lines are the model-based estimates with the shaded grey areas representing  95 percent uncertainty intervals for the reconstruction period.}\label{fig:unnamed-chunk-3}
\end{figure}

In Figure 4 we make use of the probabilistic nature of the
reconstructions by using the uncertainty to identify how likely RFI was
to exceed observed minimum/maximum values (i.e., observed during the
instrumental period) in any given year during the reconstruction period.

Figure 4A shows the probability of the RFI in the reconstruction time
period being less than the observed minimum between 1889 and 2017 for
Brisbane and Fitzroy catchments. Conversely, Figure 4B shows the
probability of the RFI in the reconstruction time period being greater
than the observed maximum. 

According to the model based reconstruction
for Brisbane (reconstruction period: 1612 - 1888), the RFI is unlikely to have fallen below the minimum
observed RFI in any year (all probabilities are below 1\%). The
reconstruction also shows the RFI was unlikely to have exceeded the
observed maximum (all probabilities were less than 5\%). In the Fitzroy
catchment (reconstruction period: 1030 - 1888), the chance of RFI falling below the minimum observed RFI
exceeds 50\% in 11 years of the reconstruction period and exceeds 75\% in
2 of those years (1877 and 1888). The RFI is likely to have exceeded the
maximum observed RFI in 3 years during the reconstruction period (1177,
1257 and 1452; \textgreater{} 50\% probability).

\newpage

\begin{figure}[!h]
\includegraphics[width=0.95\linewidth]{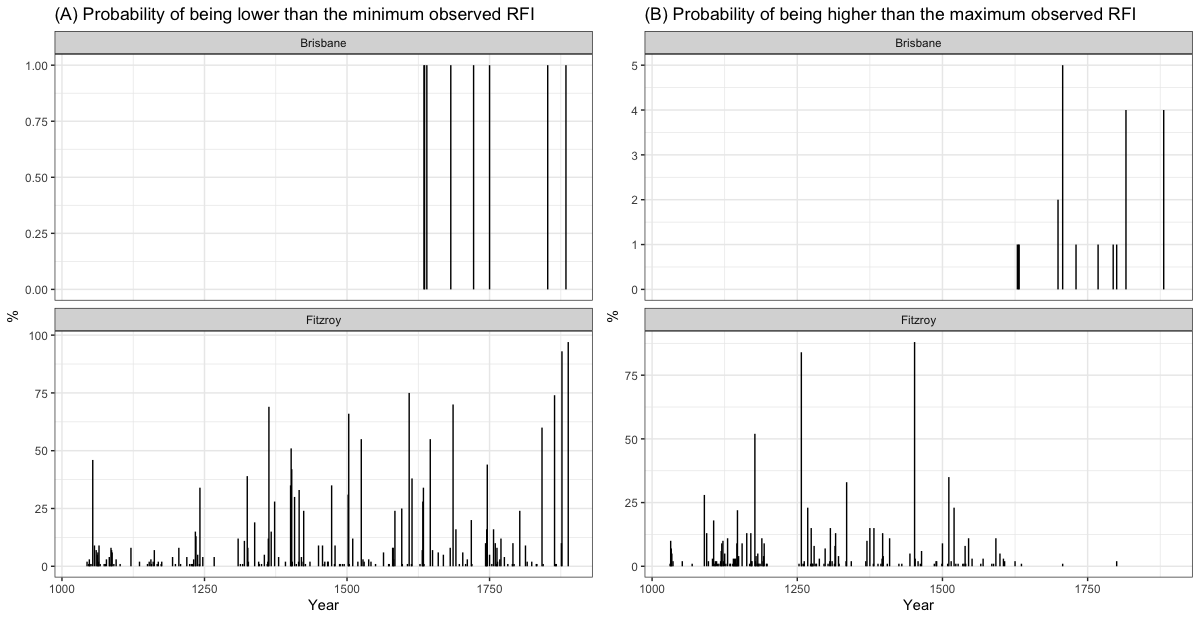} \caption{The probability of Rainfall Index (RFI) extremes for the Brisbane and Fitzroy catchment areas. The bars indicate the probability of RFI in a given reconstruction year being lower(higher) than the minimum (maximum) RFI observed in the instrumental period.}\label{fig:unnamed-chunk-4}
\end{figure}

\hypertarget{standardised-precipitation-evapotranspiration-index-reconstruction}{%
\subsection{Standardised Precipitation-Evapotranspiration Index
Reconstruction}\label{standardised-precipitation-evapotranspiration-index-reconstruction}}

Figure 5 shows SPEI (12) reconstructions for the Brisbane and Fitzroy
catchments. In Brisbane, the average estimated SPEI for the
reconstruction period is the same as the average over the instrumental
period (a value of 0.1). However, as with the RFI, the reconstruction
period estimates exhibit less variability with a standard deviation of
0.7 for the reconstruction period compared to 1.1 for the instrumental.
In Fitzroy, the average estimated SPEI is 0 over both the reconstruction
and instrumental periods. The reconstruction period estimates of SPEI
exhibit similar variability to the observed SPEI, with a standard
deviation of 1.04 compared to 1.06.

\begin{figure}[!h]
\includegraphics[width=0.9\linewidth]{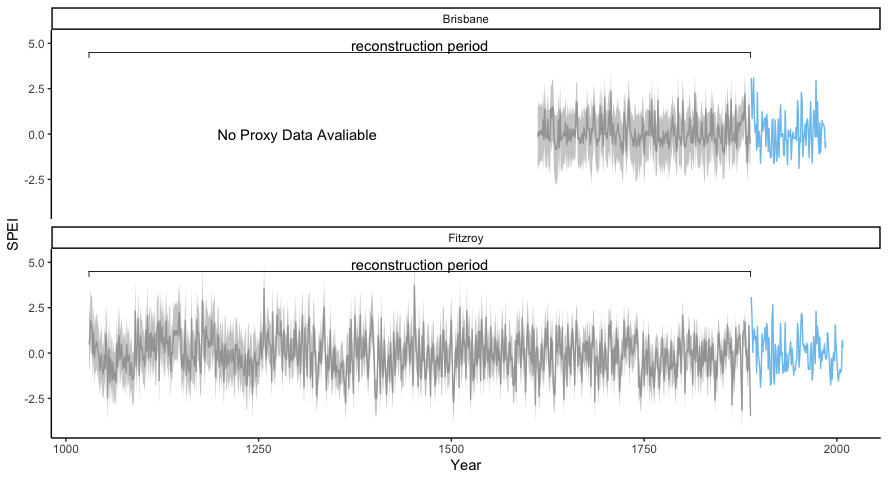} \caption{Standardised Precipitation-Evapotranspiration Index (SPEI) reconstructions for the Brisbane and Fitzroy catchment areas. The solid blue line represents the observed data. The solid grey lines are the model-based estimates for the reconstruction period and the shaded grey areas represent 95 percent uncertainty intervals for the reconstruction.}\label{fig:unnamed-chunk-5}
\end{figure}

In terms of analysing exceedance of observed minimums/maximums , Figure
6A and 6B show the probability of SPEI (12) in the reconstruction time
period being lower (higher) than the observed minimum (maximum) between
1889 and 2017 for the Brisbane and Fitzroy catchments. According to the
reconstruction for Brisbane (reconstruction period: 1612 - 1888), the chance that the SPEI fell below the
minimum or above the maximum observed SPI is below 30\% for all years.
In Fitzroy (reconstruction period: 1030 - 1888) the chance of SPEI falling below the minimum observed SPEI
exceeds 50\% in 40 years of the reconstruction period. The chance of
SPEI falling above the maximum observed SPI is greater than 50\% in only
2 years of the reconstruction period (1257 and 1452).

\begin{figure}[!h]
\includegraphics[width=0.9\linewidth]{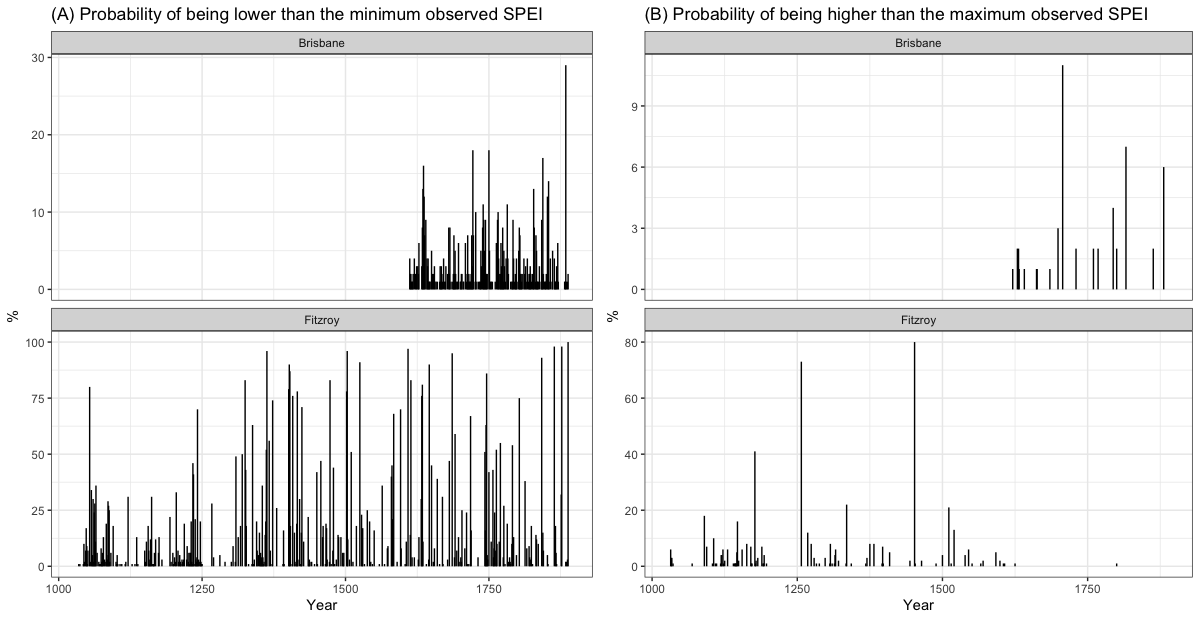} \caption{The probability of Standardised Precipitation-Evapotranspiration Index (SPEI) extremes for the Brisbane and Fitzroy catchment areas. The bars indicate the probability (expressed as a percentage) of SPEI in a given reconstruction year being lower(higher) than the minimum(maximum) SPEI observed in the instrumental period.}\label{fig:unnamed-chunk-6}
\end{figure}

\newpage
\hypertarget{discussion}{%
\section{Discussion}\label{discussion}}

Improving the management of drought, flood, water security and
hydroclimatic risk requires an understanding of long-term climate
variability. Reconstructions of rainfall, temperature and climate
divers, such as ENSO, from proxy data are not new and in recent years
there has been much attention on how such data can contribute to flood
and drought risk planning. There is a general acceptance by the water
security industry that paleoclimate proxy data has an important role to
play in future predictions of climate which is reflected in the general
uptake by industry and local governments. This reflects the general
consensus that the post-1900 instrumental period (i.e.~the period on
which all water resources infrastructure, policy, operation rules and
strategies is based) does not capture the full range of variability that
has occurred. For example, in the state of New South Wales (NSW),
regional water strategies include specific use of palaeoclimate data in
the stochastic modelling to quantify climate variability and inform on
climatic extremes such as floods and droughts (NSW Department of
Planning, Industry and Environment 2020). In particular, it uses the
high-resolution record from the Law Dome ice cores in Antarctica, which
have been used to identify eight mega-droughts (lasting from 5-39 years)
during 1000-2009 AD (Vance et al. 2015).

Until recently, palaeoclimate proxy data were often stored in
inaccessible journal papers. The construction of the PalaeoWISE database
was the first step in providing easy access to quality-assured,
standardised proxy data of relevance to Australia (Croke et al. 2021).
It also provides details on how selected proxy data correlate with the
hydroclimate indices of Queensland, allowing a user to see which proxy
data may be best to focus on in any subsequent analysis.

To date there is no standardised approach to using multiple
palaeoclimate proxy records for hydroclimate reconstructions. Previous
work has tended to use single proxy records and different statistical
techniques of varying complexity. Each method, regardless of proxy type,
will rely on a set of assumptions regarding the relationships between
proxy variables and the climate indices to be reconstructed (Birks and
John 2012). For example, the utilisation of transfer functions for
deriving palaeohydrological reconstructions from testate amoebae
assemblages is now commonplace in several regions of the world
(Krashevska et al. 2020; Schnitchen et al. 2006; Amesbury et al. 2016;
Swindles et al. 2015). Recently, diatom-based transfer function have
been used for producing quantitative palaeoclimatic reconstructions in
Southwest China (Zou et al. 2021). Principal component regression has
been used in streamflow reconstructions from tree-ring chronologies (E.
R. Cook et al. 1999; Hidalgo, Piechota, and Dracup 2001). Faraji et al.
(2021) employed principal component analysis to obtain reliable annual
hydrological variability from speleothems. Vance et al. (2013) used
spectral analysis, which allows for the study of the periodic behaviour
of time series, to explore changes in El Niño-Southern Oscillation and
rainfall variability in eastern Australia based on a 1010-yr record of
Law Dome summer sea salts (Vance 2012). Our approach adds further value
to the widespread adoption of palaeoclimate proxy data as a vital data
source on long term water planning in Australia. We note below several
advantages of our proposed approach.

\hypertarget{using-a-standardised-statistical-methodology-with-multiple-proxies}{%
\subsection{Using a standardised statistical methodology with multiple
proxies}\label{using-a-standardised-statistical-methodology-with-multiple-proxies}}

Our approach avails of the largest database of palaeoclimate proxy data
for the southern hemisphere. The database includes a total of 396
proxies that have been shown to correlate with a selected range of
hydroclimate indices across numerous geographic locations in Australia.
While we limit this paper to two case study locations, our method can be
viewed as a standard approach that can be applied at any and all
locations for which the relevant data are available. The statistical
model takes a relatively straightforward approach to quantifying the
relationship between proxies and hydroclimate indices using standard
regression and time series techniques for data calibration within the
instrumental period and reconstruction beyond the instrumental period.
The model allows for the inclusion of multiple proxies for informing the
hydroclimate reconstruction, which means that we can draw strength from
multiple data sources in an attempt to quantify the uncertainty around
past changes. The choice of what proxies to include is aided by our
filtering approach which helps us to remove proxy records that have poor
modern analogues. However, there is flexibility within the model that
still allows for the inclusion/exclusion of proxies to be user defined.
This flexibility comes with the caveat that model convergence checks
must be carried out on the results.

\hypertarget{the-advantage-of-a-bayesian-approach}{%
\subsection{The advantage of a Bayesian
approach}\label{the-advantage-of-a-bayesian-approach}}

Our model fits within a Bayesian framework which relies on the use of
MCMC algorithms to generate samples of the unknown parameters in our
statistical model. After convergence, these samples provide the
posterior distributions for any parameter in the model or any quantity
which is a function of these parameters, including, in our case,
distributions for hydroclimate reconstructions. Once we have access to
these posterior distributions we can very easily obtain point estimates,
quantiles and summaries, for example posterior medians and 95 percent
credible intervals. A major advantage of the Bayesian MCMC approach is
its flexibility. It is relatively straightforward to fit models to
complex data sets with measurement error, missing observations,
multilevel or serial correlation structures, and multiple endpoints. It
is usually much more difficult to develop the frequentist (non-Bayesian)
procedures for fitting such models. In our case, the Bayesian framework
also provides a convenient setting for inverse modelling, i.e., where we
wish to predict the value of the independent hydroclimate variable from
observations of the dependent proxy variable. We achieve this in our
model by establishing the dependence between the variables over an
instrumental time period and then placing a prior on the ``missing''
independent variables over the reconstructions period. Another notable
advantage of the Bayesian analysis approach is that the uncertainty is
handled by default within the model due to the parameters being defined
from prior probability distributions. Once the data model is combined
with the prior distributions to obtain the posteriors for the
parameters, we automatically have information on the reliability of the
estimates. The uncertainty in those estimates is propagated through any
function of the parameters which in our case includes the hydroclimate
reconstruction.

\hypertarget{limitations-of-the-approach}{%
\subsection{Limitations of the
approach}\label{limitations-of-the-approach}}

Our proposed modelling approach is not without limitations, which can be
summarised in three main points.

\begin{enumerate}
\def\labelenumi{\arabic{enumi}.}

\item This method suffers from the same issue that all reconstruction approaches do, which is that it relies on the assumption that the proxy-climate relationship remains constant in time. This is a ubiquitous problem with proxies, as the proxy variability will not solely be controlled by the climate indicator of interest. However, this problem is mitigated somewhat in our method since multiple proxy sources are used in the hydroclimate reconstruction and thus the resulting hydroclimate estimates will be informed by multiple proxy-hydroclimate relationships, not just a single proxy-hydroclimate relationship. 

\item
  The proxy filtering approach uses a threshold value of 3.5 which is user defined. This value was chosen to strike a balance between removing too many proxies and keeping proxies that cause modern analogue issues which can result in model convergence problems.  We would like to note that results may be sensitive to the choice of threshold. If users wish to be more conservative in terms of proxy removal, the filtering value can be increased, and this might decrease the number of proxies that are excluded from the model. Note, that the change may lead to model convergence problems, in which case the 3.5 threshold is recommended. 
  
\item
  The method does not provide an automatic convergence diagnosis.
  Convergence diagnostic measures must be checked by the user on a case
  by case basis. MCMC settings for the number of iterations may need to
  be adjusted in cases where convergence is not achieved, which will
  increase the computation time.
  
\item
  The quadratic component of the model could be viewed as simplistic and
  non-parametric regression approaches might allow for more flexibility.
  However, the potential increase in prediction accuracy that might be
  obtained with the increase in flexibility is not enough to warrant the
  substantial computational burden that would make the use of this model
  more difficult in a practical sense.
  
\item 
 This model does not explore spatial dependencies in the hydroclimate indices and it does not allow for assumptions that individual proxy variation might be controlled by multiple hydroclimate indices at the same time. These would be realistic modelling assumption to make, and we see future work as extending this method to account for these more complex assumptions. However, this model was specifically developed for efficiency and as a first step to giving users the ability to reconstruct independent climate indices using multiple proxy sources. 
 
\end{enumerate}

\hypertarget{potential-applications-to-the-water-security-industry}{%
\subsection{Potential applications to the water security
industry}\label{potential-applications-to-the-water-security-industry}}

The direct cost of restricted water availability to Australian urban and
rural business is estimated to be as high as \$A15 billion per year
during droughts (or \textasciitilde1.6\% of Australia's GDP; Dept.
Agriculture and Environment (2021)). Water restrictions in major cities
during the millennium drought cost \textasciitilde\$A815 million per
year and affected more than 80\% of Australian households (Dept.
Agriculture and Environment 2021). Following amendments to the
Queensland Water Act in 2018, water related climate change effects on
water availability, water use practices and the risk to land or water
resources arising from use of water on land must be considered in the
preparation of water plans. The Act also requires best practice science
to be used.

There is an increasing need to understand the full range of natural
climatic variability, particularly on decadal and longer timescales
(Head et al. 2014) and questions inevitably arise as to whether resource
assessments that are based on the recorded recent past are robust. In
the absence of direct measurements, reconstructions such as those
presented in our case study section can ultimately be used to reassess
current or baseline flood/drought risks and quantify what should be
considered ``normal'' and ``extreme'' based on longer-term past
variability.

It has been common, for example, in water resource assessments to
calculate the ``historical no-failure yield'' which essentially tests
the system's performance against the worst drought in recorded history.
However, important questions arise as to whether there were likely to
have been drought events prior to the recorded data period that are
worse and can we quantify the uncertainty around such events. As we have
demonstrated, reconstructions derived from our methodology could be used
to answer such questions and ultimately improve water resource
assessments through highlighting exposures to extreme events worse than
have been recorded and through the appropriate quantification of
uncertainties related to estimating such exposures.

According to our model based reconstructions, in Brisbane the RFI is
unlikely (probabilities between 0 and 5\%) to have exhibited extremes
beyond the minimum/maximum of what has been observed between 1889 and
2017. However, in Fitzroy there are a number of years during the
reconstruction period where the RFI is likely (\textgreater50\%
probability) to have exhibited behaviour beyond the minimum/maximum of
what has been observed. For SPEI, the probability of observing extremes
beyond the minimum/maximum of what has been observed since 1889 doesn't
exceed 50\% in any reconstruction year in the Brisbane catchment and exceeds 50\% in 2 years for the Fitzroy catchment.

Identifying solutions for hydroclimatic risk adaptation strategies that
are both optimal and robust in the presence of uncertainty presents a
difficult challenge. Reconstructions are a critical foundational data
from which a range of derivates can be extracted. Of specific relevance
to water resource modelling applications, the annual based
reconstructions can be disaggregated into daily data for model input.
Additionally, the time series can readily be classified according to
different criteria to identify events of interest for example, which
were the 10 driest years, decades, or centuries. This highlights the
versatility of catchment scale climate reconstructions to help answer a
wide range of questions that will help improve our understanding of the
longer-term context within which known historical (i.e., instrumental)
climate exists as well as a point of comparison with future projections
of climate.

The next step is to make these reconstructions available to the water
industry to incorporate into water resource modelling. For example, SEQ
has previously experienced water scarcity, when the main storage
reservoir at Wivenhoe Dam fell below 15\% during the Australian
Millennium Drought. The longer palaeoclimate reconstructions produced
here can now be used assess how often a Millennium Drought might occur
using existing rainfall-runoff derived model outputs. This is an
approach also proposed by other states such as New South Wales (NSW)
where the Department of Planning, Industry and Environment is reviewing
the useability of this palaeoclimate data for town water security
assessments under the Safe and Secure Water program (NSW Department of
Planning, Industry and Environment 2020).

It is important to stress that all reconstructions must be used in a
manner suitable for the end-user. For water resource management, the
reconstruction's time series form the input data to existing
hydrological models. In other instances, the reconstructions can be used
to re-calculate risk or to determine notable changes in climate patterns
over this longer centennial timescale. The methodology outlined in this
paper is the first to avail of all proxy data to help reconstruct
climate indices of relevance to the water industry and to help make
these freely available for operational scenarios of relevance to
critical business decisions.

\newpage

\textbf{References}

\bigskip

\hypertarget{refs}{}
\begin{CSLReferences}{1}{0}
\leavevmode\hypertarget{ref-Adams2017}{}%
Adams, J. 2017. {``{climate indices, an open source Python library
providing reference implementations of commonly used climate
indices.}''}

\leavevmode\hypertarget{ref-Amesbury2016}{}%
Amesbury, M J., G T. Swindles, A Bobrov, D J. Charman, J Holden, M
Lamentowicz, G Mallon, et al. 2016. {``{Development of a new
pan-European testate amoeba transfer function for reconstructing
peatland palaeohydrology}.''} \emph{Quaternary Science Reviews} 152:
132--51.
https://doi.org/\url{https://doi.org/10.1016/j.quascirev.2016.09.024}.

\leavevmode\hypertarget{ref-Armstrong2020}{}%
Armstrong, M S., A S. Kiem, and T R. Vance. 2020. {``{Comparing
instrumental, palaeoclimate, and projected rainfall data: Implications
for water resources management and hydrological modelling}.''}
\emph{Journal of Hydrology: Regional Studies} 31: 100728.
https://doi.org/\url{https://doi.org/10.1016/j.ejrh.2020.100728}.

\leavevmode\hypertarget{ref-Birks2012}{}%
Birks, B, and H John. 2012. {``{Strengths and Weaknesses of Quantitative
Climate Reconstructions based on Late-Quaternary Biological Proxies}.''}
\emph{Quaternary International} 279-280: 52.
https://doi.org/\url{https://doi.org/10.1016/j.quaint.2012.07.228}.

\leavevmode\hypertarget{ref-AustralianGovernment2017}{}%
Bureau of Meteorology. 2017. {``{KNOWN FLOODS IN THE BRISBANE \& BREMER
RIVER BASIN}.''} Australian Government.
\url{http://www.bom.gov.au/qld/flood/fld_history/brisbane_history.shtml}.

\leavevmode\hypertarget{ref-Cahill2016}{}%
Cahill, N, A C. Kemp, B P. Horton, and A C. Parnell. 2016. {``{A
Bayesian hierarchical model for reconstructing relative sea level: From
raw data to rates of change}.''} \emph{Climate of the Past} 12 (2).
\url{https://doi.org/10.5194/cp-12-525-2016}.

\leavevmode\hypertarget{ref-CookB2016}{}%
Cook, B I., J G. Palmer, E R. Cook, C S. M. Turney, K Allen, P Fenwick,
A O'Donnell, et al. 2016. {``{The paleoclimate context and future
trajectory of extreme summer hydroclimate in eastern Australia}.''}
\emph{Journal of Geophysical Research: Atmospheres} 121 (21): 12,
812--20, 838.
https://doi.org/\url{https://doi.org/10.1002/2016JD024892}.

\leavevmode\hypertarget{ref-Cook2002}{}%
Cook, E R., R D. D'Arrigo, and M E. Mann. 2002. {``{A Well-Verified,
Multiproxy Reconstruction of the Winter North Atlantic Oscillation Index
since a.d. 1400}.''} \emph{Journal of Climate} 15 (13): 1754--64.
\url{https://doi.org/10.1175/1520-0442(2002)015\%3C1754:AWVMRO\%3E2.0.CO;2}.

\leavevmode\hypertarget{ref-Cook1999a}{}%
Cook, E R., D M. Meko, D W. Stahle, and M K. Cleaveland. 1999.
{``{Drought Reconstructions for the Continental United States}.''}
\emph{Journal of Climate} 12 (4): 1145--62.
\url{https://doi.org/10.1175/1520-0442(1999)012\%3C1145:DRFTCU\%3E2.0.CO;2}.

\leavevmode\hypertarget{ref-Croke2021}{}%
Croke, J, J Vítkovský, K Hughes, M Campbell, S Amirnezhad-Mozhdehi, A
Parnell, N Cahill, and R Dalla Pozza. 2021. {``{A palaeoclimate proxy
database for water security planning in Queensland Australia}.''}
\emph{Scientific Data} 8 (1): 292.
\url{https://doi.org/10.1038/s41597-021-01074-8}.

\leavevmode\hypertarget{ref-DeptAgriculture2021}{}%
Dept. Agriculture, Water, and the Environment. 2021. {``{Drought
policy}.''} Australian Governement.
\url{https://www.awe.gov.au/agriculture-land/farm-food-drought/drought/drought-policy}.

\leavevmode\hypertarget{ref-Dixon2019}{}%
Dixon, B C., J T. Tyler, B J. Henley, and R Drysdale. 2019. {``{Regional
patterns of hydroclimate variability in southeastern Australia over the
past 1200 years}.''} \emph{Earth and Space Science Open Archive}.
https://doi.org/\url{https://doi.org/10.1002/essoar.10501482.1}.

\leavevmode\hypertarget{ref-Dowdy2019}{}%
Dowdy, A J., A Pepler, A Di Luca, L Cavicchia, G Mills, J P. Evans, S
Louis, K L. McInnes, and K Walsh. 2019. {``{Review of Australian east
coast low pressure systems and associated extremes}.''} \emph{Climate
Dynamics} 53 (7): 4887--4910.
\url{https://doi.org/10.1007/s00382-019-04836-8}.

\leavevmode\hypertarget{ref-Faraji2021}{}%
Faraji, M, A Borsato, S Frisia, J C. Hellstrom, A Lorrey, A Hartland, A
Greig, and D P. Mattey. 2021. {``{Accurate dating of stalagmites from
low seasonal contrast tropical Pacific climate using Sr 2D maps, fabrics
and annual hydrological cycles}.''} \emph{Scientific Reports} 11 (1):
2178. \url{https://doi.org/10.1038/s41598-021-81941-x}.

\leavevmode\hypertarget{ref-Head2014}{}%
Head, L, M Adams, H V. McGregor, and S Toole. 2014. {``{Climate change
and Australia}.''} \emph{WIREs Climate Change} 5 (2): 175--97.
https://doi.org/\url{https://doi.org/10.1002/wcc.255}.

\leavevmode\hypertarget{ref-Herbert2016}{}%
Herbert, A V., and S P. Harrison. 2016. {``{Evaluation of a
modern-analogue methodology for reconstructing Australian palaeoclimate
from pollen}.''} \emph{Review of Palaeobotany and Palynology} 226:
65--77.
https://doi.org/\url{https://doi.org/10.1016/j.revpalbo.2015.12.006}.

\leavevmode\hypertarget{ref-Hernandez2020}{}%
Hernández, A, G Sánchez-López, S Pla-Rabes, L Comas-Bru, A Parnell, N
Cahill, A Geyer, R M. Trigo, and S Giralt. 2020. {``{A 2,000-year
Bayesian NAO reconstruction from the Iberian Peninsula}.''}
\emph{Scientific Reports} 10 (1): 14961.
\url{https://doi.org/10.1038/s41598-020-71372-5}.

\leavevmode\hypertarget{ref-Hidalgo2001}{}%
Hidalgo, H G., T C. Piechota, and J A. Dracup. 2001. {``{Alternative
principal components regression procedures for dendrohydrologic
reconstructions}.''} \emph{Surface Water and Climate} 36 (11).

\leavevmode\hypertarget{ref-Ho2015}{}%
Ho, M, A S. Kiem, and D C. Verdon-Kidd. 2015. {``{A paleoclimate
rainfall reconstruction in the Murray-Darling Basin (MDB), Australia: 1.
Evaluation of different paleoclimate archives, rainfall networks, and
reconstruction techniques}.''} \emph{Water Resource Research} 51 (10).

\leavevmode\vadjust pre{\hypertarget{ref-Hyndman2021}{}}%
Hyndman, R, G Athanasopoulos, and M O'Hara-Wild. 2021. {``{fpp3:
Forecasting: Principles and Practice}.''}

\leavevmode\hypertarget{ref-Braganza2009}{}%
K, Braganza, J L. Gergis, S B. Power, J S. Risbey, and Fowler A M. 2009.
{``{A multiproxy index of the El Ni{ñ}o--Southern Oscillation, A.D.
1525--1982}.''} \emph{Journal of Geophysical Research: Atmospheres} 114
(D5). https://doi.org/\url{https://doi.org/10.1029/2008JD010896}.

\leavevmode\hypertarget{ref-Kiem2020}{}%
Kiem, A S., T R. Vance, C R. Tozer, J L. Roberts, R. Dalla Pozza, J.
Vitkovsky, K Smolders, and M A. J. Curran. 2020. {``{Learning from the
past -- Using palaeoclimate data to better understand and manage drought
in South East Queensland (SEQ), Australia}.''} \emph{Journal of
Hydrology: Regional Studies} 29: 100686.
https://doi.org/\url{https://doi.org/10.1016/j.ejrh.2020.100686}.

\leavevmode\hypertarget{ref-Krashevska2020}{}%
Krashevska, V, A N. Tsyganov, A S. Esaulov, Y A. Mazei, K A. Hapsari, A
Saad, S Sabiham, H Behling, and S Biagioni. 2020. {``{Testate Amoeba
Species- and Trait-Based Transfer Functions for Reconstruction of
Hydrological Regime in Tropical Peatland of Central Sumatra, Indonesia
}.''} \url{https://www.frontiersin.org/article/10.3389/fevo.2020.00225}.

\leavevmode\hypertarget{ref-Luterbacher2001}{}%
Luterbacher, J, E Xoplaki, D Dietrich, P D. Jones, T D. Davies, D
Portis, J F. Gonzalez-Rouco, et al. 2001. {``{Extending North Atlantic
oscillation reconstructions back to 1500}.''} \emph{Atmospheric Science
Letters} 2 (1-4): 114--24.
https://doi.org/\url{https://doi.org/10.1006/asle.2002.0047}.

\leavevmode\hypertarget{ref-McKee1993}{}%
McKee, T B., N J. Doesken, and J Kleist. 1993. {``{The relationship of
drought frequency and duration to time scales.}''} In \emph{Eighth
Conference on Applied Climatology, American Meteorological Society}.

\leavevmode\hypertarget{ref-Michel2020}{}%
Michel, S, D Swingedouw, M Chavent, P Ortega, J Mignot, and M Khodri.
2020. {``{Reconstructing climatic modes of variability from proxy
records using ClimIndRec version 1.0}.''} \emph{Geosci. Model Dev.} 13
(2): 841--58. \url{https://doi.org/10.5194/gmd-13-841-2020}.

\leavevmode\hypertarget{ref-NSWDOP2020}{}%
NSW Department of Planning, Industry and Environment. 2020. {``{Regional
water strategies in New South Wales}.''} New South Wales.
\url{https://www.industry.nsw.gov.au/water/plans-programs/regional-water-strategies}.

\leavevmode\hypertarget{ref-ODonnell2021}{}%
O'Donnell, A J., W L. McCaw, E R. Cook, and P F. Grierson. 2021.
{``{Megadroughts and pluvials in southwest Australia: 1350--2017 CE}.''}
\emph{Climate Dynamics}.
\url{https://doi.org/10.1007/s00382-021-05782-0}.

\leavevmode\vadjust pre{\hypertarget{ref-OHara2021}{}}%
O'Hara-Wild, M, R Hyndman, E Wang, G Caceres, TG Hensel, and T Hyndman.
2021. {``{fable: Forecasting Models for Tidy Time Series}.''}

\leavevmode\hypertarget{ref-Parnell2015}{}%
Parnell, A C., J Sweeney, T K. Doan, M Salter-Townshend, J R. M. Allen,
B Huntley, and J Haslett. 2015. {``{Bayesian inference for palaeoclimate
with time uncertainty and stochastic volatility}.''} \emph{Journal of
the Royal Statistical Society: Series C (Applied Statistics)} 64 (1):
115--38. https://doi.org/\url{https://doi.org/10.1111/rssc.12065}.

\leavevmode\hypertarget{ref-Plummer2003}{}%
Plummer, M. 2003. {``{JAGS: A program for analysis of Bayesian graphical
models using Gibbs sampling}.''} \url{https://mcmc-jags.sourceforge.io}.

\leavevmode\vadjust pre{\hypertarget{ref-Plummer2019}{}}%
Plummer, M, S Alexey, and M Denwood. 2021. \emph{{rjags: Bayesian
Graphical Models using MCMC}}. R package version 4-12.

\leavevmode\vadjust pre{\hypertarget{ref-MASS}{}}%
Ripley, B D., W N. Venables, K Hornik, A Gebhardt, and D Firth. 2021.
\emph{MASS: Modern Applied Statistics with s}. R package version 7.3-54.
\url{http://www.stats.ox.ac.uk/pub/MASS4/}.

\leavevmode\hypertarget{ref-Schnitchen2006}{}%
Schnitchen, C, D J. Charman, E Magyari, M Braun, I Grigorszky, B
Tóthmérész, M Molnár, and Zs. Szántó. 2006. {``{Reconstructing
hydrological variability from testate amoebae analysis in Carpathian
peatlands}.''} \emph{Journal of Paleolimnology} 36 (1): 1--17.
\url{https://doi.org/10.1007/s10933-006-0001-y}.

\leavevmode\hypertarget{ref-Group2013}{}%
Suncorp Group. 2013. {``{Risky Business: Insurance and Natural Disaster
Risk Management.}''} Queensland.
\url{https://www.google.com/url?sa=t\&rct=j\&q=\&esrc=s\&source=web\&cd=\&ved=2ahUKEwjg8bKshM3yAhUeRUEAHWmRCZoQFnoECAIQAQ\&url=https\%3A\%2F\%2Fwww.aph.gov.au\%2Fparliamentary_business\%2Fcommittees\%2Fhouse_of_representatives_committees\%3Furl\%3Djscna\%2Fsubs\%2Fsub0151-\%252}.

\leavevmode\hypertarget{ref-Swindles2015}{}%
Swindles, G T., J Holden, C L. Raby, T E. Turner, A Blundell, D J.
Charman, M W. Menberu, and B Kløve. 2015. {``{Testing peatland
water-table depth transfer functions using high-resolution hydrological
monitoring data}.''} \emph{Quaternary Science Reviews} 120: 107--17.
https://doi.org/\url{https://doi.org/10.1016/j.quascirev.2015.04.019}.

\leavevmode\vadjust pre{\hypertarget{ref-Lasso}{}}%
Tibshirani, Robert. 1996. {``Regression Shrinkage and Selection via the
Lasso.''} \emph{Journal of the Royal Statistical Society. Series B
(Methodological)} 58 (1): 267--88.
\url{http://www.jstor.org/stable/2346178}.

\leavevmode\hypertarget{ref-Tingstad2014}{}%
Tingstad, A H., D G. Groves, and R J. Lempert. 2014. {``{Paleoclimate
Scenarios to Inform Decision Making in Water Resource Management:
Example from Southern California's Inland Empire}.''} \emph{Journal of
Water Resources Planning and Management} 140 (10): 4014025.
\url{https://doi.org/10.1061/(ASCE)WR.1943-5452.0000403}.

\leavevmode\hypertarget{ref-Tozer2016}{}%
Tozer, C R., T R. Vance, J L. Roberts, A S. Kiem, M A. J. Curran, and A.
D. Moy. 2016. {``{An ice core derived 1013-year catchment-scale annual
rainfall reconstruction in subtropical eastern Australia}.''}
\emph{Hydrol. Earth Syst. Sci.} 20 (5): 1703--17.
\url{https://doi.org/10.5194/hess-20-1703-2016}.

\leavevmode\hypertarget{ref-Vance2012}{}%
Vance, T R. 2012. {``{Annualized Summer Sea Salt From the Law Dome Ice
Core Chemistry Record, 1000-2009, Ver. 1}.''} \emph{Australian Antarctic
Data Centre}.
\href{https://doi:10.26179/5d50ef2192dc4}{doi:10.26179/5d50ef2192dc4}.

\leavevmode\hypertarget{ref-Vance2013}{}%
Vance, T R., T D. van Ommen, M A. J. Curran, C T. Plummer, and A D. Moy.
2013. {``{A Millennial Proxy Record of ENSO and Eastern Australian
Rainfall from the Law Dome Ice Core, East Antarctica}.''} \emph{Journal
of Climate} 26 (3): 710--25.
\url{https://doi.org/10.1175/JCLI-D-12-00003.1}.

\leavevmode\hypertarget{ref-Vance2015}{}%
Vance, T R., J L. Roberts, C T. Plummer, A S. Kiem, and van Ommen T D.
2015. {``{Interdecadal Pacific variability and eastern Australian
megadroughts over the last millennium}.''} \emph{Geophysical Research
Letters} 42 (1): 129--37.
https://doi.org/\url{https://doi.org/10.1002/2014GL062447}.

\leavevmode\hypertarget{ref-Vehtari2021}{}%
Vehtari, A, A Gelman, D Simpson, B Carpenter, and P C. Bürkner. 2021.
{``{Rank-Normalization, Folding, and Localization: An Improved
\(\widehat{R}\) for Assessing Convergence of MCMC (with Discussion)}.''}
\emph{Bayesian Analysis} 16 (2): 667--718.
\url{https://doi.org/10.1214/20-BA1221}.

\leavevmode\hypertarget{ref-Zou2021}{}%
Zou, Y, L Wang, H He, G Liu, J Zhang, Y Yan, Z Gu, and H Zheng. 2021.
{``{Application of a Diatom Transfer Function to Quantitative
Paleoclimatic Reconstruction --- A Case Study of Yunlong Lake, Southwest
China }.''}
\url{https://www.frontiersin.org/article/10.3389/feart.2021.700194}.

\end{CSLReferences}

\bibliographystyle{unsrt}
\bibliography{references.bib}

\newpage

\footnotesize \textbf{Data References from Table 2}

\begin{enumerate}
\def\labelenumi{\alph{enumi}.}
\item
  Lough, J.M. Northeast Queensland 350 Year Summer Rainfall
  Reconstructions, NOAA National Centers for Environmental Information,
  2011. \url{https://www.ncdc.noaa.gov/paleo/study/10292}
\item
  Palmer, J.G. Palmer - Waihora Terrace - PHTR - ITRDB NEWZ058, NOAA
  National Centers for Environmental Information, 2002.
  \url{https://www.ncdc.noaa.gov/paleo-search/study/4083}
\item
  Hendy, E.J., Gagan, M.K., Lough, J.M. Kurrimine Beach, Brook Island,
  Britomart Reef, Great Palm Island, Lodestone Reef, Pandora Reef,
  Havannah Island - Luminescence master chronology, NOAA National
  Centers for Environmental Information, 2003.
  \url{https://www.ncdc.noaa.gov/paleo-search/study/1918}
\item
  Xiong, L., Palmer, J.G. Xiong - Werberforce - LIBI - ITRDB NEWZ075,
  NOAA National Centers for Environmental Information, 2002.
  \url{https://www.ncdc.noaa.gov/paleo-search/study/5378}
\item
  Buckley, B.M., Anchukaitis, K.J., Cook, B.I., Canh Nam, Le. Buckley -
  Bidoup Nui Ba National Park - FOHO - ITRDB VIET001 NOAA National
  Centers for Environmental Information, 2010.
  \url{https://www.ncdc.noaa.gov/paleo-search/study/10453}
\item
  Urban, F.E., Cole, J.E., Overpeck, J.T. Maiana - Data, NOAA National
  Centers for Environmental Information, 2000.
  \url{https://www.ncdc.noaa.gov/paleo-search/study/1859}
\end{enumerate}

\newpage
\begin{table}
 \caption{Validation results. Summaries of cross-validation results based on standardized data are shown for  hydroclimate indices in the test data set. Labels in the Climate Variable column are consistent with the PaleoWise database. The results are only presented for records that passed convergence checks and are ordered based on coverage (highest to lowest). SPEI = Standardised Precipitation-Evapotranspiration Index, SPI = Standardised Precipitation Index. Coverage is defined as the percentage of time that the true value is captured with the 95\% uncertainty intervals. The mean error is the average prediction error. RMSE is the Root Mean Squared Error. N is the number of datapoints. *Number in brackets indicates the aggregation period in months.}
  \centering
  \begin{tabular}{lp{.25\textwidth}p{.13\textwidth}p{0.09\textwidth}p{0.07\textwidth}p{0.07\textwidth}}
    \toprule
    \multicolumn{2}{c}{}  \\
    \cmidrule(r){1-2}
    Climate variable  & Description*  & Coverage \par (target 95\%) & Mean \par Error & RMSE & N \\
    \midrule
    {\bf All} &  & {\bf 90} & {\bf -0.06} & {\bf 1.04} & {\bf 930}    \\
    \hline
     rain\_1yr\_wy7  & Average annual \par rainfall  & 93 &  -0.16 & 1.05 &  180 \\
     avg\_temp\_1yr\_wy7 & Average  annual \par temperature & 91 & 0.11 & 0.99 & 150 \\
     spi\_12m\_1yr\_wy7  & SPI  extreme flood \par index (12)    & 91    & -0.13 & 1.09 & 150 \\
     spei\_12m\_1yr\_wy7 & SPEI extreme drought  \par index (12) & 90 &  0.01 &  1.09    & 150 \\
     spei\_36m\_1yr\_wy7  & SPEI extreme drought \par  index (36) & 90  & -0.09 & 0.92 & 30  \\
     et\_morton\_potential\_1yr\_wy7 & Evapotranspiration \par (Morton method) & 89 & -0.33 & 1.08 & 90 \\
     spei\_24m\_1yr\_wy7 & SPEI  extreme drought \par index (24) & 87 &  0.12   & 1.04 &    120\\
     spei\_48m\_1yr\_wy7 & SPEI  extreme drought \par index (48) & 87 &  -0.05   & 0.84 &    60\\

     \bottomrule
  \end{tabular}
  \label{tab:table}
\end{table}

\newpage
\begin{table}
 \caption{Data Summary. A summary of the available proxy data related to the Rainfall Index (RFI) and the Standardised Precipitation-Evapotranspiration Index for a 12-month aggregation period (SPEI (12)) for the Brisbane and Fitzroy catchments. Columns 3 and 4 shows the number of proxy datasets before and after the filtering was applied. Columns 5 and 6 show the range of years where the proxy records begin and end. Column 7 shows the start and end points of the instrumental time period. Column 8 references the dataset IDs as they appear in the PalaeoWISE Database.}
  \centering
  \begin{tabular}{p{.1\textwidth}p{.11\textwidth}p{.12\textwidth}p{0.12\textwidth}p{0.12\textwidth}p{0.12\textwidth}p{0.12\textwidth}p{0.13\textwidth} }
    \toprule
    \multicolumn{2}{c}{}  \\
    \cmidrule(r){1-2}
    Catchment & Hydroclimate \par Index&    \# proxy \par records \par (before filter) &    \# proxy \par records \par (after filter)   & range \par start-years \par (after filter)&   range  \par end-years  \par (after filter) &    instrumental \par period \par start-end  &  PalaeoWISE \par Dataset ID \par (after filter) \\
    \hline
Brisbane &  RFI & 9  & $6^{a-c}$ & 1984-2001    & 1876-1612 & 2001-1889 &$11^{a}$, $2^{a}$, \par $224^{b}$, $269^{c}$, \par $8^{a}$, $9^{a}$ \\
Fitzroy & RFI   & 22 & $18^{a,c,d,e}$ & 1983-2008 & 1881-1030   & 2008-1889 & $10^{a}$, $11^{a}$, \par  $13^{a}$, $14^{a}$, \par $15^{a}$, $16^{a}$, \par $17^{a}$, $2^{a}$,\par $205^{d}$,$245^{e}$, \par $246^{e}$, $269^{a}$, \par $3^{a}$, $4^{a}$, \par $5^{a}$, $6^{a}$, \par $8^{a}$, $9^{a}$\\
Brisbane &  SPEI (12)   & 13 &  $9^{a-c}$ &
1981-2001   & 1876-1612 &   2001-1889   & $11^{a}$, $13^{a}$, \par $15^{a}$, $2^{a}$, \par $224^{b}$, $269^{c}$, \par $4^{25}$, $8^{a}$, $9^{a}$\\
Fitzroy & SPEI (12) & 21 &  $18^{a,c,e,f}$ & 1981-2008 &    1881-1030 & 2008-1989    & $10^{a}$, $11^{a}$, \par $13^{a}$, $14^{a}$, \par $15^{a}$, $16^{a}$, \par $168^{f}$, $17^{a}$,\par $2^{a}$, $245^{e}$,\par  $246^{e}$, $269^{c}$, \par $3^{a}$, $4^{a}$, \par $5^{a}$, $6^{a}$, \par $8^{a}$, $9^{a}$  \\
 \bottomrule
  \end{tabular}
  \label{tab:table}
\end{table}

\end{document}